\documentclass[12pt]{iopart}

\usepackage{epsfig}
\usepackage{graphicx}
\usepackage{dcolumn}
\usepackage{bm} 
\newcommand{\beq}{\begin{equation}}
\newcommand{\eeq}{\end{equation}}
\newcommand{\beqa}{\begin{eqnarray}}
\newcommand{\eeqa}{\end{eqnarray}}

\newcommand{\om}{\omega}

\def\pra#1{{ Phys.\ Rev. A\/} {\bf#1}}

\def\prl#1{{ Phys.\ Rev.\ Lett.} {\bf#1}}

\begin{document}

\title{Elliptical Trajectories in Nonsequential Double Ionization}

\author{Xu Wang and J. H. Eberly}

\address{Rochester Theory Center and the Department of Physics 
\& Astronomy\\
University of Rochester, Rochester, New York 14627}
\ead{wangxu@pas.rochester.edu}
\begin{abstract}
Using a classical ensemble method, nonsequential double ionization is predicted to exist with elliptical and circular polarization. Recollision is found to be the underlying mechanism and it is only possible via elliptical trajectories.

\end{abstract}


\maketitle

\section{Introduction}
An anomalously high degree of electron correlation is well known to be present in multiphoton atomic double ionization (see reviews \cite{Agostini-DiMauro, Becker-Rottke}). Two ionization channels are available: the atom may either lose the two electrons one by one, which is called sequential double ionization (SDI), or lose the two electrons at the same time, which is called nonsequential double ionization (NSDI).  Compared to SDI, which can be well explained by the ADK tunneling theory \cite{ADK} based on a single active electron approximation, NSDI is a much more complicated (and thus interesting) phenomenon of atomic physics. Recently, open questions in this domain associated with departures from the usual linear polarization of the laser pulses have received close theoretical attention \cite{Shvetsov-Shilovski-etal, Wang-Eberly09, circ-chaos}.

\begin{figure}[b!]
\hspace{2cm}
\includegraphics[width=8cm]{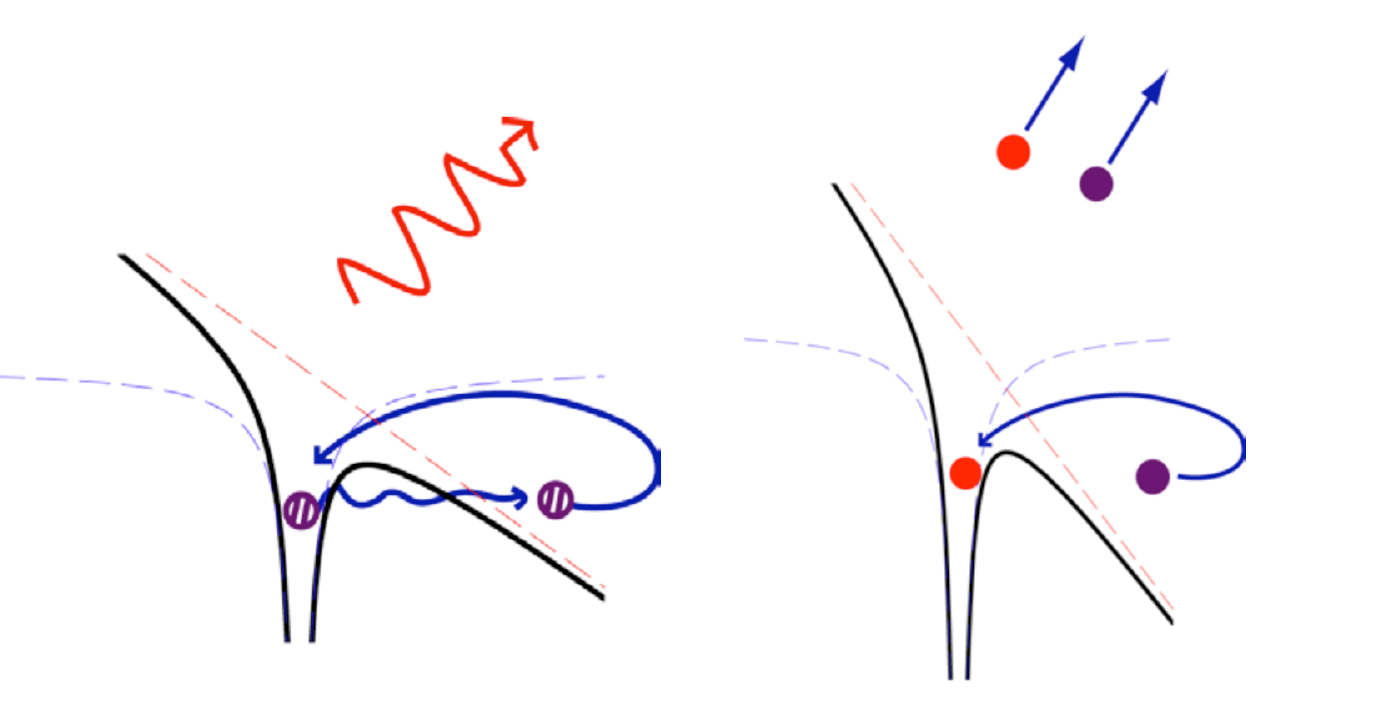}
\caption{{ \footnotesize \label{f.pptrecolln} High harmonic generation and double ionization via the three-step scenario, in which the reversed laser phase directs the first ionized electron back to the core. Linear polarization of the laser pulse is implied, in order that the returning electron collide with the nucleus  (see \cite{Schafer-etal, Corkum}).}}
\end{figure}

The accepted scenario for such high 2e correlation is a three-step  model in which one electron is ionized via tunneling when the Coulomb barrier is tipped down by the laser electric field. Second, the electron is carried away and then driven back to the ion core as the laser field reverses phase. Third, the energized returning electron recombines at the nucleus to generate high harmonics \cite{Schafer-etal, Corkum}, or kicks out a second electron \cite{Corkum} to provide double ionization, as sketched in Fig. \ref{f.pptrecolln}.

\begin{figure}[h!]
\hspace{4cm}
\includegraphics[width=5cm]{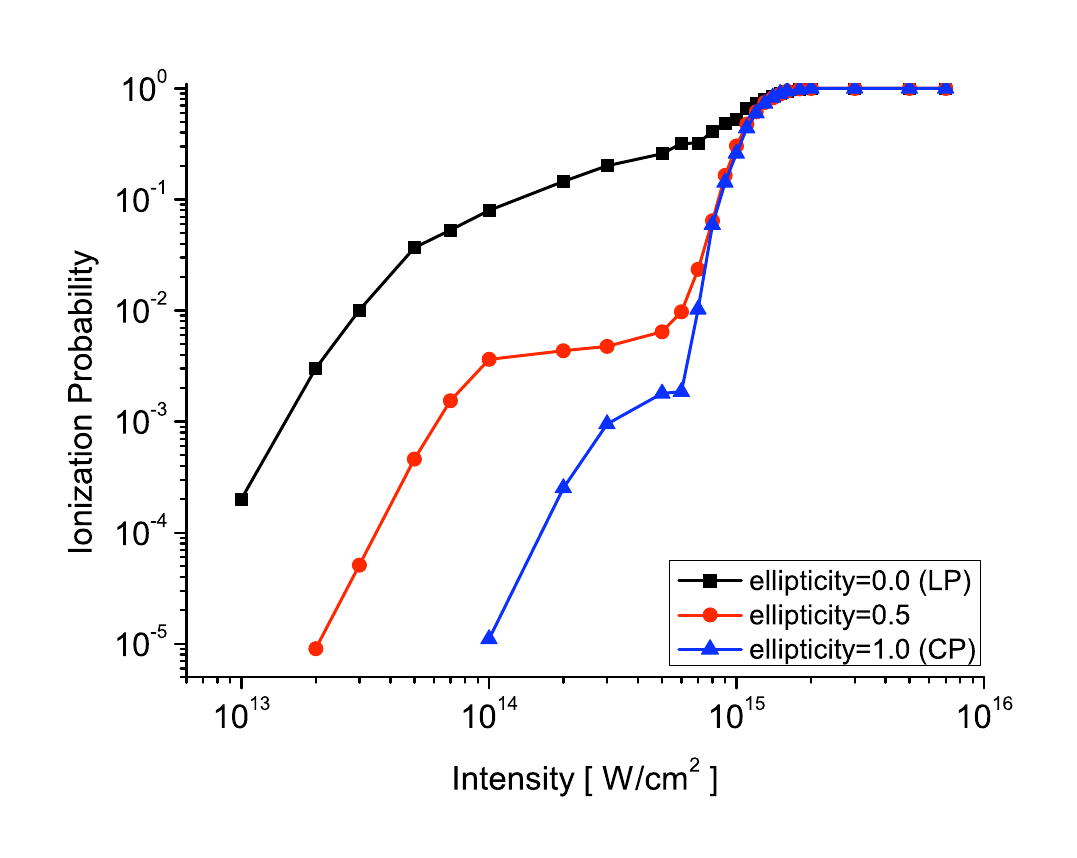}
\caption{{ \footnotesize \label{f.knee} Ionization probability obtained in numerical experiments as a function of laser peak intensity under different ellipticities for a model ``Xe/Kr" atom. So-called knee structure is evident for a full range of ellipticities.}}
\end{figure}

This classically-inspired picture leads to the understanding that little or no NSDI can be expected if the laser field is elliptically polarized, since the first electron cannot return to the ion core under elliptical polarization. This is at odds with some experiments. Characteristic NSDI events have been observed with the molecules NO and O$_2$ \cite{Guo-Gibson} and atomic magnesium \cite{Gillen-etal} under circular polarization. We report here an explanation for these phenomena, and show why they do not require invention of a different non-collision scenario. To be clear that we are dealing with NSDI as it is usually understood, we show in Fig. \ref{f.knee} that the characteristic knee signature for NSDI is present for all degrees of ellipticity, up to and including circular polarization.

\section{Elliptical trajectories}

\begin{figure}[b!]
\includegraphics[width=4.5cm]{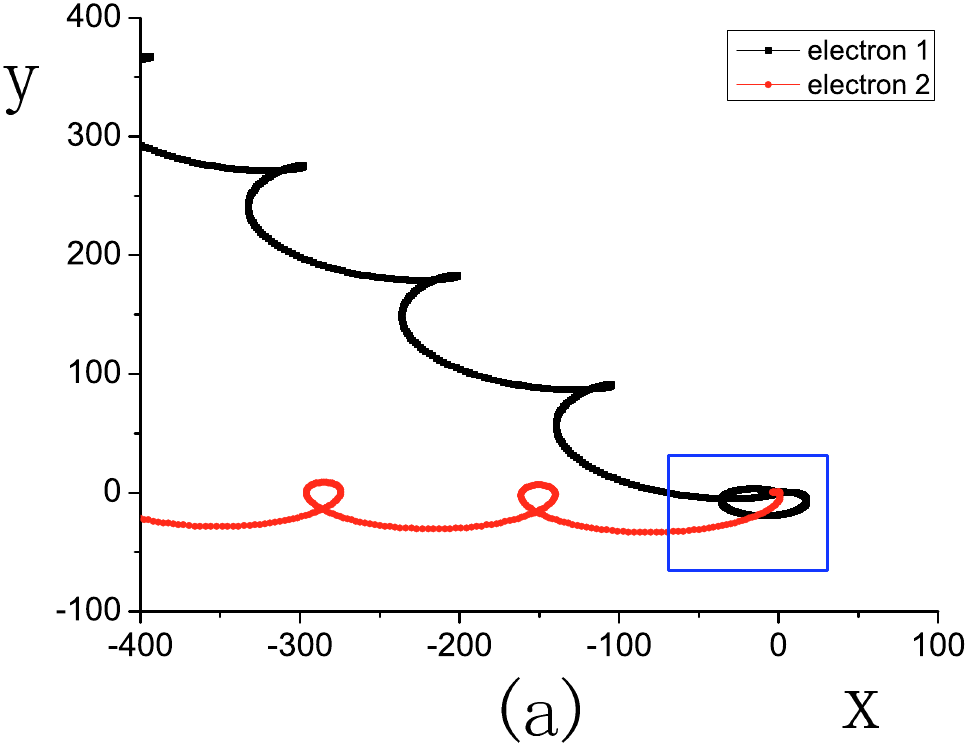}
\hspace{3cm}
\includegraphics[width=4.5cm]{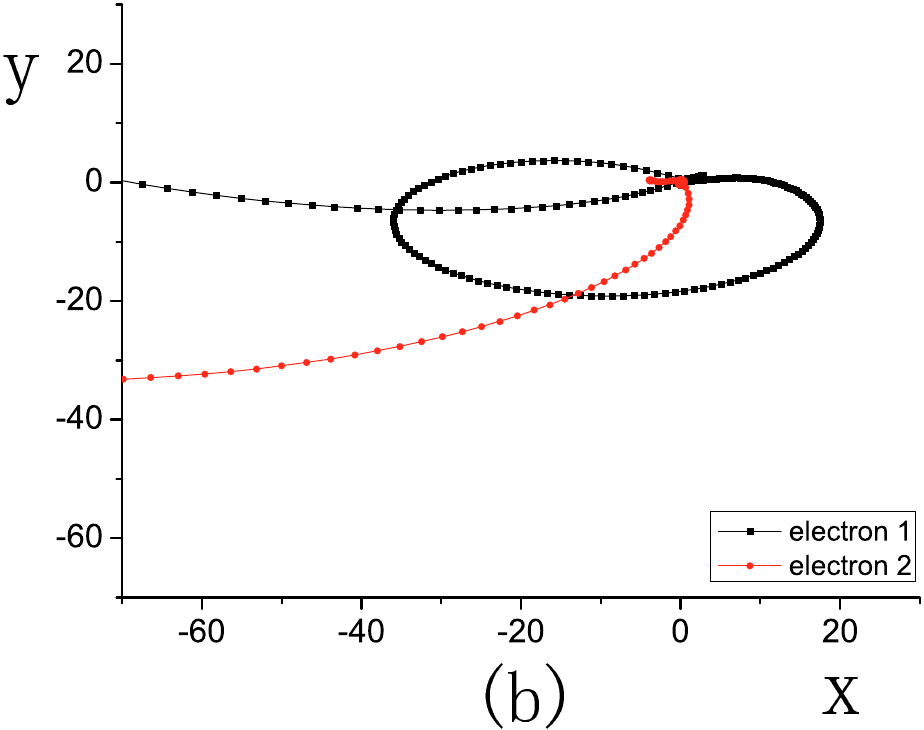} \\ \\
\includegraphics[width=4.5cm]{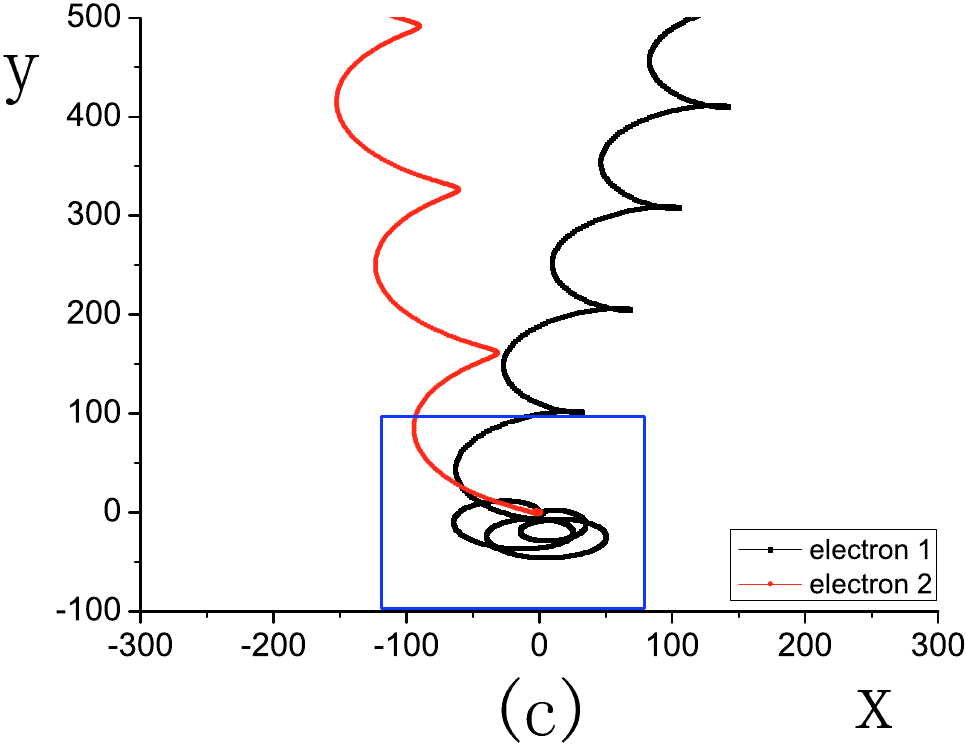}
\hspace{3cm}
\includegraphics[width=4.5cm]{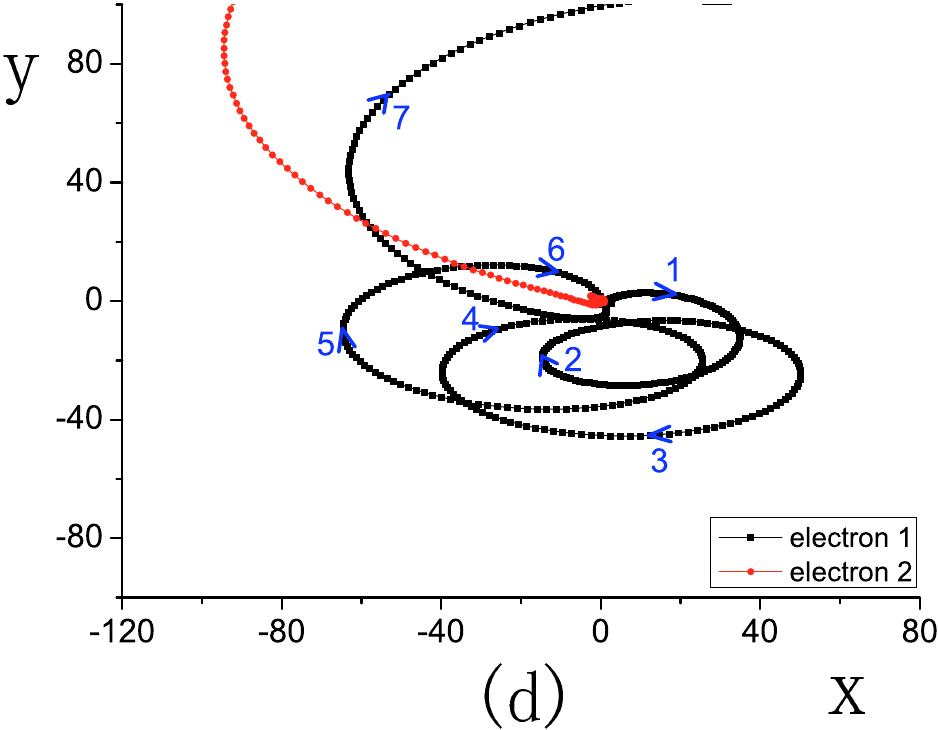} 
\caption{{ \footnotesize \label{f.recollision} Two typical recollision trajectories for ellipticity 0.5. The blue boxes of the left two figures are greatly expanded to give the right two figures. Top: the recolliding electron's trajectory is an almost perfect ellipse. It returns to the origin in one laser cycle and kicks out the second electron. Bottom: the elliptical trajectory of the recolliding electron is distorted by the ion's Coulomb force and executes several cycles before coming close enough to the ion core to kick out the second electron. The arrows and numbers are used to show the temporal motion of the first electron. }}
\end{figure}

To understand NSDI with elliptical polarization we have undertaken a large series of numerical ``experiments"  (see last section below for details), in which two electrons are followed along classical trajectories. Successful NSDI events were back-analyzed to determine the sequence of events that lead to an NSDI conclusion. Back-analysis of classical trajectories is a very powerful tool to investigate detailed physical processes \cite{Panfili-etal02} and a quantum analog using wave function masking has also been developed by Haan, et al. \cite{Haan-etal02}. 

The surprising conclusion of this back analysis is that every successful NSDI trajectory is elliptical! That is, in all cases examined, the first ionized electron circulates along an elliptical trajectory for one or more cycles and then collides and ejects the second electron. Typical trajectories showing one-cycle and multi-cycle ellipses are given in Fig. \ref{f.recollision}. Thus the answer to the question raised by the apparent failure of the recollision scenario for polarizations that are not linear is that there is no failure -- recollision is still the mechanism. More than that, recollision is possible {\em only via elliptical trajectories}.

\section{Understanding the elliptical trajectories}

\begin{figure}[h!]
\hspace{3cm}
\includegraphics[width=6cm,height=2.5cm]{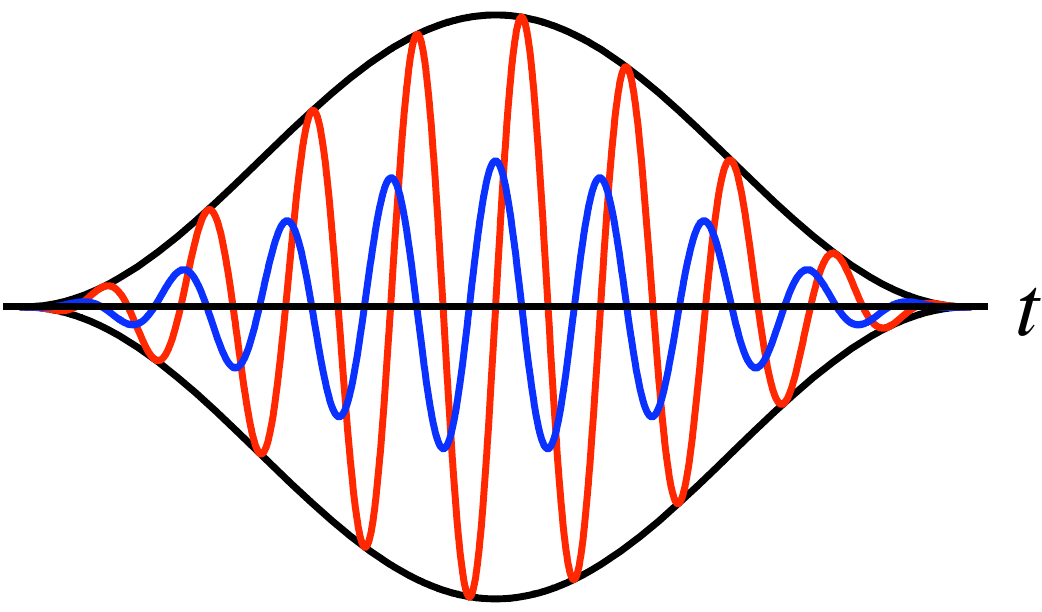}
\caption{{ \footnotesize \label{f.pulse} Illustration of an elliptically polarized laser pulse. The red curve is the electric field along the x direction and the blue curve is the electric field along the y direction. The ellipticity is 0.5 here.}}
\end{figure}

The dominance of the elliptical trajectories can be understood from a simple analysis based on free electron kinematics. Our analysis begins with the common assumption that electron motion after ionization is adequately described by free electron physics in the presence of the laser field. The core's Coulomb field plays a role, but it is secondary, so the force felt by an electron comes mainly from the elliptically polarized field 
\beq 
\vec{E}(t) = E_0 f(t)[\hat{x}\sin(\om t + \phi) +  \varepsilon \hat{y} \cos(\om t + \phi)],
\eeq 
as sketched in Fig. \ref{f.pulse}. Here $\phi$ is a random phase from shot to shot in our series of numerical laser-atom interactions, and $f(t)$ is a slowly varying several-cycle envelope whose time dependence can be ignored in first approximation. 

The charge of the electron is -1 in atomic units so the free-electron velocities along the x and y directions are (having put $f$ = 1):
\beqa
v_x(t) &=& v_{0x} + \int_{t_0}^t a_x (t') dt' \nonumber\\
&=& v_{0x} + \frac{1}{\om} E_0 [\cos(\om t + \phi) - \cos(\om t_0 + \phi)], \label{e.vx}
\eeqa
and
\beqa
v_y(t) &=& v_{0y} - \frac{\varepsilon}{\om} E_0 [\sin(\om t + \phi) - \sin(\om t_0 + \phi)]. \label{e.vy}
\eeqa
The first terms are the initial velocities at the time of ionization, independent of the laser force. By the nature of first escape over the tipped Coulomb barrier (or via tunneling through it), the initial electron velocity in the polarization direction is zero. So $v_{0x} = 0$. However, the initial velocity in the transverse direction can be nonzero. The exact form of the transverse distribution is not important, and has only been derived in a one-electron model \cite{Goreslavsky-Popruzhenko}, but its existence is important and has begun to be taken into account (see a recent treatment of ionization under elliptical polarization by Shvetsov-Shilovski, et al. \cite{Shvetsov-Shilovski-etal}). The second terms are the electron oscillation velocities in the field and the last terms are laser-induced drift velocities. 

Because the ionization is strongly dependent on field strength the first electron is most probably ionized at the peak of the field strength. To a first approximation, for ellipticities not close to circular, this makes $\sin(\om t_0 + \phi) \approx \pm1$ and $\cos(\om t_0 + \phi) \approx 0$. 
Under this scenario the electron has the drift velocity $v_{yd}$ in the y direction:
\beq
v_{yd} = v_{0y} \pm  \frac{\varepsilon E_0}{\om}.
\eeq 
This drift velocity has been used to obtain the first formula for NSDI probability as a function of ellipticity \cite{Wang-Eberly10}. Apart from this drift motion, one sees that the trajectory is a perfect ellipse, which we call the ``circulation ellipse", with half major axis $E_0/ \om^2$ and half minor axis $\varepsilon E_0/ \om^2$.    

For intensity $6 \times 10^{14}$ W/cm$^2$ and wave length 780 nm, we have $E_0/ \om^2 \approx$ 40 a.u. so the field-dependent part of the transverse drift distance for one laser cycle is about $(\varepsilon E_0 / \om) \times (2\pi / \om) = 2 \pi \varepsilon E_0 / \om^2$, which is about 6 times larger than the size of the circulation ellipse. This means that the electron will quickly be taken away from the ion core and no recollision could occur. However, from the range of velocities $v_{0y}$ that are available, there will be a compensating value, allowing a perfect elliptical oscillation that returns the first electron to the ion core, for a collision that may kick out the second electron.

\section{The role of the ion core Coulomb potential}   

The ion core Coulomb potential, which is neglected in the free electron analysis of the previous section, plays a role in increasing the probability of NSDI. As explained above, balancing the laser-induced drift velocity along the y direction with a compensating transverse initial ionization velocity is the reason for successful recollisions. Exact balancing yields a purely elliptical trajectory, and a nearly perfect example is shown in Fig. \ref{f.recollision} (a) and (b). If the balancing is not so precise, the first ionized electron will drift away, although the drifting velocity is small. With the aid of the ion core Coulomb potential, however, slow drifts will be pulled back. Slowly drifting circulation ellipses can be pulled back and forth around the ion core several times before the first electron finds the second one and kicks it out. This is shown in Fig. \ref{f.recollision} (c) and (d). In fact, we find in our numerical experiments that most of the successful recollision trajectories fall into this category.   

\section{Method}

The method that we have used to obtain the numerical experiment data presented is the classical ensemble method, which has been described in detail in \cite{ClassicalEnsemble}. The contribution of the classical ensemble method to interpretation of double ionization phenomena is well known and has been critically analyzed \cite{classicalNSDI, Rudenko-etal}. A microcanonical but otherwise essentially random phase-space-filling distribution of initial conditions (typically 10$^6$-10$^7$ in number) is generated using a many-pilot-atom method \cite{Abrines} before turning on the laser field. 

The classical atom model has a $Z = +2$ charged ion core fixed at the origin, and the energy of the two-electron bound system is given by
\beq
E_{tot} = \frac{p_1^2}{2} + \frac{p_2^2}{2} - Z V_{sc}(\vec r_1, a) - Z V_{sc}(\vec r_2, a) + V_{sc} (\vec r_1 - \vec r_2, b), \nonumber   
\eeq
with $V_{sc}$ standing for the soft-core Coulomb potential of the familiar ``Rochester" form \cite{Softcore}:
\beq
V_{sc} (\vec r, c) = \frac{1}{\sqrt{r^2+c^2}}. \nonumber
\eeq     
The energy of each 2e member of the ensemble is set to be -1.3 a.u., which is close to the binding energies of both Kr (-1.41 a.u.) and Xe (-1.23 a.u.). (The binding energy of a 2e system is calculated as the negative sum of the 1st and 2nd ionization potentials. For example, the 1st and 2nd ionization potentials of Kr are 13.999 eV and 24.359 eV, which yields a binding energy of -(13.999+24.359) eV = -38.358 eV = -1.41 a.u.)

An elliptically polarized pulse as shown in Fig. \ref{f.pulse} is then applied to the system and at each moment the positions and momenta of both electrons are recorded, as determined by solution of the relevant coupled Newtonian equations. Experience with this method \cite{Ho-Eberly3e} has shown that out-of-plane effects can be neglected under current experimental conditions so we only work on the x-y plane.

\section{Summary}
In summary, we have shown that the characteristic NSDI knee structure still appears with elliptical and circular polarization. This is consistent with results of highly cited experiments \cite{Guo-Gibson, Gillen-etal}. Elliptical polarization leads to elliptical trajectories and our numerical experiments show that all successful NSDI events come from recollisions at the end of an elliptical trajectory. The dynamical process determining these recolliding elliptical trajectories is explained using free electron kinematics. The residual effect of Coulomb attraction to the nucleus plays a secondary role in many cases.

\section{Acknowledgement}
Partial financial support was provided by DOE Grant DE-FG02-05ER15713.

\end{document}